\documentclass[pre,preprint]{revtex4}
 \usepackage{graphics}
 \begin{document}
 \draft
 \bibliographystyle{revtex4}
 \titlepage
 \title{
The effect of synchronized area on SOC behavior in a kind
of Neural Network Model
 \footnote{The project supported by P.R.China
 National Basic Research Project "Nonlinear Science"}}

\author{Xiao Wei Zhao and Tian Lun Chen}
 \email{xwzhao@263.net  chenzh@nankai.edu.cn}
  \affiliation{Department of Physics, Nankai University,
   Tianjin 300071, P.R.China}
 \date{}
 \begin{abstract}
Based on the LISSOM model and the OFC earthquake model, we introduce a
self-organized feature map Neural Network model . It
displays a "Self Organized Criticality"(SOC) behavior. It
can be seen  that the feature area (synchronized
area)produced by self-organized process brings about some
definite effect on SOC behavior and the system evolves into
a "partly-synchronized" state. For explaining this
phenomena , a quasi-OFC earthquake model is simulated.
 \end{abstract}
 \maketitle
 %\keywords{
 %neuron network , self-organized criticality , avalanche ,
 %partly-synchronized}

\section{Introduction}

A few years ago, Bak, Tang and Wiesenfeld introduced the
concept of the "Self-Organized Criticality" (SOC) in sand
pile model~\cite{1}. From then on, this concept has been
widely studied in some extended dissipative dynamical
systems, such as earthquake ~\cite{2}, biology evolution
~\cite{3}, and so on. It is shown that all these systems
can naturally evolve into a "critical state" with no
intrinsic spatial and temporal scales through a
self-organized process without the need to fine-tune
parameters of the system. This critical state is
characterized by a power-law distribution of avalanche
sizes, where the size is the total number of toppling
events or unstable units.

    Now,the research on SOC has come into a new level. One
has studied several factors' influence on SOC behavior,
such as network size, periodic or nonperiodic boundary
conditions,local dynamics variable is conservative or not,
and so on. Many investigators believe that it is
intrinsic-stability(order) and variability(disorder)'s
common action make the system evolves into a "frozen
disorder" (SOC) state~\cite{4,5,6}. It is the combination
feature of stability and variability, and it's complex
spatial-temporal dynamic behavior, make the system in SOC
state have maximum complexity and latent computing potency.

    The brain is a complex system and its
information process has the properties of stability and
variability -- on one hand, there are relative stable
information stored mechanism, and stored area in brain(such
as feature area in cortex) ; on the other hand, the brain
is influenced by the environment and one should continuous
update knowledge and concepts. The similarity between the
SOC systems and the brain has lead us to study Artifical
Neural Network(ANN) and SOC together. There is some SOC
behavior shown in the neuron network model introduced by
our group~\cite{7}.

    The brain is also a complex system with highly
complexity, highly order and special structure. The
structure must have the effect on the brain's dynamics
behavior. Our neuron network can also produce some special
structure, so we believe that the SOC behavior shown in our
model must have it's own special behavior and rule.

    In this paper, the feature area(produced by
self-organized process)'s definite effect on SOC behavior
has been studied.

\section{Model}

Here we propose a two-dimensional neural network model of
square lattice. This model is a kind of serial
self-organized neural network model, based on the LISSOM
model ~\cite{8}. When a  $h$- dimensional vector $\zeta$ is
inputted, the state $\eta_{ij}(t)$  of the neuron $(i,j)$
at time $t$  is changed according to the formula:
\begin{eqnarray}
  \eta_{ij} &=& \sigma\{ \sum\limits_h \mu_{ij,h} \zeta_h  +\gamma_e \sum_{kl}
                 E_{ij,kl}\eta_{kl}(t-1)   \nonumber  \\
   \mbox{} & \mbox{} &
       - \gamma_i \sum \limits_{k^\prime l^\prime}
I_{ij, k^\prime l^\prime} \eta_{k^\prime l^\prime}(t-1) \}
\nonumber
\\
       & = &  \sigma\{ f_{ij}(t) \}
\end{eqnarray}
where $\sigma(x)$  is an active function, and we design it
as sign function, i.e., if $x \ge 0$, then  $\sigma(x)=1$ ,
otherwise $\sigma(x)=-1$ . $\mu_{ij,h}$  is an afferent
input weight vector; $E_{ij,kl}$  is the excitatory lateral
connection weight on the connection from the neuron $(k,l)$
to the $(i,j)$ neuron; $I(ij,kl)$  is the inhibitory
connection weight. $f_{ij}(t)$ is the local field of the
neuron $(i,j)$  at time $t$. $\gamma_e$ and $\gamma_i$ are
constant factors. The adjustment of those three connection
weights is as following according to the dynamic Hebb rule:
\begin{eqnarray}
\nonumber & & \mu_{ij,h}(t+1)=\frac{\mu_{ij,h}(t)+\alpha
\eta_{ij} \xi_h} {\{
\sum_h\left[\mu_{ij,h}(t)+\alpha\eta_{ij}\xi_h
\right]^2\}^{1/2}}
 \mbox{,}
  \\
 \nonumber
& & E_{ij,kl}(t+1)=\frac{E_{ij,h}(t)+\alpha_E \eta_{ij}
\eta_{kl}} { \sum_{ij} \left[E_{ij,kl}(t)+\alpha_E
\eta_{ij} \eta_{kl}\right]}
 \mbox{,}
 \\
 & &
 I_{ij,k^\prime l^\prime }(t+1)=\frac{I_{ij,k^\prime l^\prime}(t)+
\alpha_I \eta_{ij} \eta_{k^\prime l^\prime}} {
\sum_{k^\prime l^\prime}\left[I_{ij,{k^\prime
l^\prime}}(t)+ \alpha_I \eta_{ij} \eta_{k^\prime l^\prime}
\right]}  \mbox{.}
\end{eqnarray}
where $\alpha, \alpha_E, \alpha_I$ are the learning rates.

Note the change of the neuron states is quick and the
adjustment of the connection weights is slow. Usually,
after over 10 iterations of the neuron states when any
pattern is inputted (at this time, the network state
becomes an attractor in the state space, often a fixed
point.), all connection weights are updated once. Thus,
after learning a while, the lateral connection weight
self-evolves into the"Mexican hat" profile,the afferent
input weight self-organizes into a topological map of the
input space, the neuron network can produce some special
feature areas, and the state of the neural network is
evolved from disordered case to stable and topological case
in state space.

Then we introduce the following interactive process between
the neurons, similar with the pulse coupled interaction:

1) When the neuron $(i,j)$  is stable, i.e.,
$\eta_{ij}(t)=\sigma\{f_{ij}(t-1)\}$ , it doesn't influence
the others;

2) If the neuron $(i,j)$  is unstable, i.e.,  $\eta_{ij}
\ne \sigma\{f_{ij}(t-1)\}$ or $f_{ij}(t-1)=0$ , then the
nearest neighbors $(i^\prime,j^\prime)$  around this
unstable neuron will receive a pulse respectively and their
local fields will be changed. At the same time, the neuron
$(i,j)$
  becomes stable again, depending on the formula (1);

3) When all neurons of the neural network are stable, we
choose the minimum $g$ among the absolute values of all
local fields $f_{ij}$ and drive the local field of every
neuron, i.e.,
\begin{equation}
f_{ij}\rightarrow f_{ij}-c * \eta_{ij} * g
\end{equation}
where  $c$ is a constant.

Now, we present the computer simulation procedure of this
model in detail:

1) Variable initialization.   In the 2-dimensional  $n
\times n$ neural network model, let the initiatory state
and local field equal $0$; random initialize each
connection weight among $\left[-1,1\right]$ ; and produce
$M$ random input patterns, $\zeta_j^i \in [-1,1]$,
$(i=1,2,\cdots M; j=1,2,\cdots h)$.

2) Learning process.   According to formula (1), we input
the pattern circularly and iterate the neuron state and
local field. After period of time, the space state of the
network reaches stability and we consider the $M$ input
patterns have been stored.

3) Associative memory.   Input a new pattern, and then
search the unstable neuron $(i,j)$   as defined above in
whole neural network. Due to being unstable, the neuron
$(i,j)$ discharges a pulse to the each nearest neighbor
$(i^\prime , j^\prime)$ and thus causes the local fields of
them to change as following:
\begin{equation}
f_{i^\prime j^\prime}\rightarrow f_{i^\prime
j^\prime}-\frac{\gamma}{2} \eta_{i^\prime j^\prime}
(1+|f_{ij}|)
\end{equation}
where $\gamma$  represents the pulse intensity, symbol $|\
|$ denotes absolute value.

Simultaneously, according to formula (1), the neuron
$(i,j)$ becomes stable as: $\eta_{ij}\rightarrow \sigma
(f_{ij})$ , $f_{ij}\rightarrow \eta_{ij}$  ; where
$\sigma(x)$  is sign function.

Repeat this procedure until all neurons of the model are
stable. Define one avalanche as all unstable neurons in
this process. Then begin drive process by formula (3) and
new avalanche.

\section{Simulation results}

Recently, Bak and Sneppen have investigated  the power law
distribution $P(X)$ of the distances $X$ between subsequent
unstable sites in lattice of BS biology evolution
model~\cite{3}, and J.De.Boer et al. have found some
interested result with it~\cite{9}.So in this paper we
studied not only the distribution $P(S)$ of the avalanche
sizes $S$ but also the distribution $P(X)$ of the distances
$X$ between the subsequent unstable sites . We find the
distribution will deviate from the power law in some
conditions.

\subsection{The effect of synchronized area on SOC
behavior}

The size of our lattice is $ 40\times40$.We find that in
associative memory process, the distribution of the
avalanche sizes has power-law behavior,
 $P(S) \propto S^{-\tau}$  , $\tau \approx 0.90$ .
It is shown in Fig.1.a. The distribution $P(X)$  of the
distances $X$ between the subsequent unstable sites has
power-law behavior too, $P(X) \propto X^{-\beta}$ , $\beta
\approx 2.23$ , it is shown in Fig.1.b.

The SOC behavior changed with the scope of lateral
connection has been studied. We increase the excitatory
lateral connection radius $d_e$ and the inhibitory
connection radius $d_i=3d_e$. It can be seen in Fig 1.a
that avalanche size and occurring probability of large
scale avalanche  decrease with the slope $d_e$ increasing,
and the distribution of $P(X)$ deviates from power law more
and more in large $X$, the probability of large distance
$X$ between the subsequent unstable sites also increases,
we consider it is a deviation from SOC behavior, it is
shown in Fig 1.b.

By investigating , we think  our system is in a
"partly-synchronized" state, hence the dynamics behavior
mentioned above can be seen.

A.Corral  et al. propose SOC state and synchronization
state might be considered as two uttermost state of system
(just like two sides of the same coin)~\cite{4}. The
inhomogeneity introduced by boundary or initialization
conditions can propagate into interior of network, hence
makes the system evolve into SOC state~\cite{10,11}. When
inhomogeneity is not large enough, the system finds a
compromise between synchronization and SOC~\cite{6}. It can
be considered as a partly-synchronized state.

If Our Model is only a pure OFC model without learning
process, the system will present a macroscopic SOC behavior
among almost all the lattices, but it is also a neuron
network model. As a kind of self-organized feature map
model, after learning a while, it's neurons will develop a
unique lateral interaction "Mexican hat" profile that
represents its long-term associations with each other. the
afferent input weights will self-organize into a
topological map of the input space~\cite{8}, it can form
some special topological feature regions. We consider that
in these regions, as the connection weights become more
topographically ordered, neuron's synchronization effect
between each other will be reinforced. When  order in these
regions is applied to the model, the system has a tendency
from  SOC state to synchronized state. At last, the system
finds a compromise between synchronization and SOC, it
could be seen as "partly-synchronized" state. With the
process, the distribution of avalanche varies from the
continuous distribution to a discrete one, the possibility
of large scale avalanche propagating into the interior of
the synchronization regions will reduce, and occurring
probability of large scale avalanche will decrease too, the
one-off isolated avalanche(only one unstable site in an
avalanche)in these regions  will increase greatly, it makes
the distances $X$ between the subsequent unstable sites
have a stochastic spatial even distribution in the area.
This distribution has more effect on probability of large
distance between unstable sites than probability of small
distance. (Because probability of small distance is larger
than one of large distance.)

So at this moment, there are some areas in synchronized
state and another regions in SOC state. We can consider
that with the increasing of $d_e$, the feature
region(synchronized region) produced by self-organized
process becomes wider, which results in the whole system
dynamics  deviating largely from SOC state, it can be seen
in Fig.1.

To verify the idea mentioned above, we draw the avalanche's
distribution map of the whole system. We draw  one
avalanche's distribution snapshot every 1000 avalanches,
then overlap all the snapshots in one picture. The result
can be seen in Fig2. It can be clearly seen that the blank
region (seldom avalanche area) expands with the increasing
of $d_e$. It means that the synchronized region introduced
by self-organized expands and large scale avalanches reduce
more and more. Even though, there are still some isolated
unstable neurons in the area, it indirectly verifies  our
deduction of $p(X)$ distribution mentioned above.

We investigate the relation between average avalanche size
$\langle S \rangle$ and radius $d_e$. From Fig.3 , we can
see that with the decreasing of $d_e$ , $\langle S \rangle$
will increases, and with $d_e$ approaching 0, the slope of
the curve becomes quite large. This  result is approximate
to the phenomena with the increasing of pulse discharging
intensity $\gamma$~\cite{7}, and is consistent with the
result in Ref.~\cite{12}. It implies that the network
approaches to SOC state when $d_e$ approaching 0 or
$\gamma$ approaching 0.5.

When the ratio $T$ of  radius $d_i$ with radius $d_e$ is
changed , the probability distribution $P(S)$ and $P(X)$
and avalanche distribution are changed too, the tendency is
similar to Fig.1, Fig.2. With the increment of $T$, the
partly-synchronized behavior of system becomes
distinctness. The phenomena can also be explained by the
expanding of synchronized region , but it leads us to think
that the inhibitory lateral connection and excitatory
lateral connection  have what different effect on
synchronized process? It is look like that the inhibitory
lateral connection has more important effect, but it is not
very clear, there are still a lot of work to do.

 \subsection{The approximate behavior in a kind of quasi-OFC earthquake
model}

OFC earthquake model is a kind of SOC model which has been
widely studied in recent years~\cite{2}.If we use
transformation in our model : $1-\eta_{ij}
f_{ij}\rightarrow F_{ij}$ , then formula (4) become
$F_{i^\prime j^\prime}\rightarrow F_{i^\prime
j^\prime}+\frac{\gamma}{2}F_{ij}$ . It is the avalanche
mechanism of the OFC earthquake model , in fact these two
models belong to the same class.Therefore we add some
synchronized regions in OFC model and examine that if the
system has the similarity partly-synchronized behavior as
in our neuron network model. Hence we introduced a cellular
automaton model based on OFC model, the steps are as
follows:

1) We define a $N\times N$ square lattice ,and  a  $L\times
L$ square area in the lattice ( $L=KN<N$) .

2) Initialize all sites to a random value $F_{ij}$ between
0 and 1.

3) If any $F_{ij} \geq F_{th} =1$,then redistribute the
force on $F_{ij}$ to its neighbors according to the rule:
\begin{eqnarray}
F_{i^\prime j^\prime}\rightarrow F_{i^\prime
j^\prime}+\frac{\gamma}{2} F_{ij}\ \ \ \
\
   \left\{\begin{array}{ll}
     \gamma=\alpha<0.5 & ,\ when \ site \ (i,j)  \in  L \times L \
     region,\\
     \gamma=0.5 & ,otherwise.

  \end{array}\right.
\end{eqnarray}
\ \ \ \ \ \ \ \ \ \
$
F_{ij}  \rightarrow  \ 0
$

4) Repeat step 3 until no $F_{ij} \geq F_{th}$,we define
the avalanche is fully evolved.

5) Locate the site with the largest strain ,$F_{max}$ ,then
drive all sites
\begin{equation}
F_{ij} \rightarrow F_{ij} + ( F_{th} - F_{max} )
\end{equation}
and return to step 3. Here we use open boundary conditions
,and serial working mode.It is same as our neuron network
model but different from traditional OFC model.

We still focus on the spatial distribution of $S$ and $X$.
Let $\alpha=0.1$, then change $K$, the result is shown in
Fig 4. It can be seen that the probability distribution
$P(X)$ increases and deviates from power-law more with $K$
increasing at larger $X$ in Fig 4.a. The tendency is
similar to that in our previous neuron network model. In
Fig 4.b , we can see the occurring probability of large
scale avalanche reduces with $K$ increasing too, but the
phenomena is not distinctness as that in previous model.
The reason may be that the previous model has more complex
structure than this model. It produces many pieces of
synchronized region, but here, we only add a piece of
synchronized region in model.

A. Corral, Grassberger et al. have investigated the
influence of pulse discharging intensity $\gamma$ in OFC
model. They propose the model present a macroscopic
synchronization among all the elements of the lattice when
$\gamma$ is small~\cite{4,13}. Therefore in this model ,
with expanding of $L\times L$ area($\gamma=\alpha=0.1$),
the synchronized area in the network expands, the
synchronization behavior becomes distinctness, and the
whole system evolves into a partly-synchronized state. The
dynamics behavior of the previous model is so consistent
with this model, it suggested that maybe they  have the
similar dynamics mechanism . Our results tend to agree with
this idea.

For studying the partly-synchronized phenomena further, the
relation between distribution $P(x)$ and $\alpha$ has been
investigated. Let $K=0.8$ , the partly-synchronized
behavior become distinctness with $\alpha$ decreasing from
0.5 . But when $\alpha$ less than a number (about 0.4),
there is no more distinctness  partly-synchronized behavior
introduced by $\alpha$ changing (seen in Fig 5) . It can
show that the $L\times L$ area have evolved into
synchronized state. We also draw the whole network's
avalanches distribution map(Fig .6). It can be clearly seen
that the isolated avalanches' number in $L\times L$ area
increases with $\alpha$ decreasing, and the large scale
avalanche can't propagate into the interior of $L\times L$
area. These result are consistent with the work of C.Tang ,
Grassberger et al.~\cite{10,13}.
%%%%%%%%%%%%%%%%%%%%%%%%%%%%%%%%%%%%%%%%%%%%%%%%%%%%%%%%%%%%%%%%%%%%%%
\section{conclusion}
In this paper, we analyze the dynamics of the
proposed neural network model and find the distribution of
the avalanche sizes and the distances $X$ between
subsequent unstable sites show the power-law behavior. More
important, we find the function area and weight
distribution produced by self-organized process in our
Neural Network model will let the $P(S)$ and $P(X)$
distribution deviate from power law behavior , and the
system evolves into a partly-synchronized state at this
time. To verify the explanation , we study a quasi-OFC
earthquake model containing synchronized region , and find
it will deviate from power-law , evolve into a
partly-synchronized state in some conditions too.

Our self-organized feature map Neuron Network model is just
a very simple simulation of brain. The real brain has very
complex structure and more specific feature regions in
Cortex. So brain may express a
quasi-SOC(partly-synchronized) behavior more than a pure
SOC behavior. Therefore  the stored pattern in brain might
be designed as a quasi-SOC attractor, and the associate
memory process might be designed as the process of the
input pattern evolving into the attractor.

Now the neuron synchronization in brain has been observed
in many experiments~\cite{14}. Rodriguez et al. have
investigated the long distance synchronization of human
brain activity~\cite{15}. We think it would be interesting
to further investigate the relationship between
synchronization and cognitive , associate process.
 \printfigures{}
%%%%%%%%%%%%%%%%%%%%%%%%%%%%%%%%%%%%%%%%%%%%%

 \newpage
 \begin{figure}
  %Put xwzhaofig1a.eps and xwzhaofig1b.eps in here
 \caption{The distribution $P(S)$ of the
avalanche sizes $S$, and the distribution $P(X)$ of
distances $X$ between subsequent unstable sites, changed
with lateral connection radius $d_e$ , for our $40\times
40$ neuron network model. Where $d_i=2d_e$, $\gamma=0.5$.
 (a). Log-log plot of $ P(S)$ .vs. sizes $S$.
 (b). Log-log plot of $ P(X)$ .vs. distances $X$.}
 \label{f1}
 \end{figure}
 \begin{figure}
 %Put xwzhaofig2a.eps,xwzhaofig2b.eps
 %    xwzhaofig2c.eps,xwzhaofig2d.eps in here
 \caption{
The avalanche distribution map for
our $40\times 40$ neuron network model , where $d_i=2d_e$,
$\gamma=0.5$. $(\circ)$ represent the origin sites of
avalanche. shadow(+) represent the avalanche region.
(a)-(d) correspond to $d_e=1,2,4,6$}
 \label{f2}
 \end{figure}
 \begin{figure}
  %Put xwzhaofig3.eps in here.
 \caption{
The avalanche average size $\langle S \rangle$ as a function of the
lateral connection radius $d_e$ for our $40\times 40$
neuron network model , where $d_i=2d_e$, $\gamma=0.5$. With
decreasing of $d_e$, $\langle S \rangle$ increases,and with $d_e$
approaching 0, the slope of the curve become quite large.}
 \label{f3}
 \end{figure}
 \begin{figure}
 %Put xwzhaofig4a.eps xwzhaofig4b.eps in here
 \caption{
The distribution $P(X)$ of the distance $X$ between
subsequent unstable sites and the distribution $P(S)$ of
the avalanche sizes $S$ , changed with $K=L/N$. for our
$40\times 40$ quasi-OFC earthquake model. Where in $L
\times L$ area,$\gamma=\alpha=0.1$, out of this area,
$\gamma=0.5$.
 (a). Log-log plot of $ P(X)$ .vs. distances $X$ .
 (b). Log-log plot of $ P(S)$ .vs.  sizes $S$. }
 \label{f4}
 \end{figure}
 \begin{figure}
 %Put xwzhaofig5a.eps xwzhaofig5b.eps in here
 \caption{
The distribution $P(X)$ of the distance $X$ between
subsequent unstable sites and the distribution $P(S)$ of
the avalanche sizes $S$ , changed with pulse discharging
intensity $\alpha $ in $L \times L$ area for our $40\times
40$ quasi-OFC earthquake model. Where $K=L/N =0.8 $, out of
$ L \times L $ area $\gamma=0.5$.
 (a). Log-log plot of $ P(X)$ .vs. distances $X$.
 (b). Log-log plot of $P(S)$ .vs. sizes $S$.}
 \label{f5}
 \end{figure}
 \begin{figure}
  %Put xwzhaofig6a.eps,xwzhaofig6b.eps
 %    xwzhaofig6c.eps,xwzhaofig6d.eps in here
 \caption{
The avalanche distribution map for our $40\times 40$
quasi-OFC earthquake model , where $K=L/N=0.8$, out of $L
\times L $ area $\gamma=0.5$. pulse discharing intersity
$\alpha $ is changed. $(\circ)$ represent the origin sites
of avalanche. shadow(+) represent the avalanche region. The
map(a)-(d) correspond to $\alpha=0.5,0.48,0.4,0.1$}
 \label{f6}
 \end{figure}

\newpage
\begin{thebibliography}{99}
\bibitem{1}
P. Bak, C. Tang, and K. Wiesenfield, Phys. Rev. A 38 (1988)
364
\bibitem{2}
Z. Olami, S. Feder, and K. Christensen, Phys. Rev. Lett. 68
1244 (1992)
\bibitem{3}
 P. Bak, and K. Sneppen, Phys. Rev. Lett. 71 , 4083(1993)
\bibitem{4}
 A.Corral, et al.,Phys.Rev.Lett.74 , 118(1995)
\bibitem{5}
  N.Mousseau, Phys. Rev. Lett. 77 , 968(1996)
\bibitem{6}
  S. Bottani, Phys. Rev. lett. 74 , 4189(1995)
\bibitem{7}
  Huang Jing, and Chen TianLun to be published in Comm. Theor.Phys
\bibitem{8}
  J. Sirosh, and R. Miikkulainen, Biol. Cybern. 71 , 65(1994)
\bibitem{9}
  J. De. Boer et al. Phys. Rev. E 51 , 1059(1995)
\bibitem{10}
 A.A Middleton, and C.Tang  Phys. Rev. Lett. 74 , 742(1995)
\bibitem{11}
  H.Ceva, Phys. Lett. A 245 , 413(1998)
\bibitem{12}
  M. L. Chabanol, and V. Hakim, Phys. Rev. E 56 , R2343(1997)
\bibitem{13}
 P. Grassberger, Phys. Rev. E. 49 , 2436(1994)
\bibitem{14}
  Singer W.Annul. Rev. Physiol. 55 , 359(1993);

  Singer W. and Gray C. M. Annul. Rev. Neurosci. 18 , 555(1995) ;

  Nelson J. L. et al. Vis. Neuronsci 9 , 21(1992)
\bibitem{15}
  Eugenio Rodriguez, et al. Nature 397(1999)
\end{thebibliography}
 
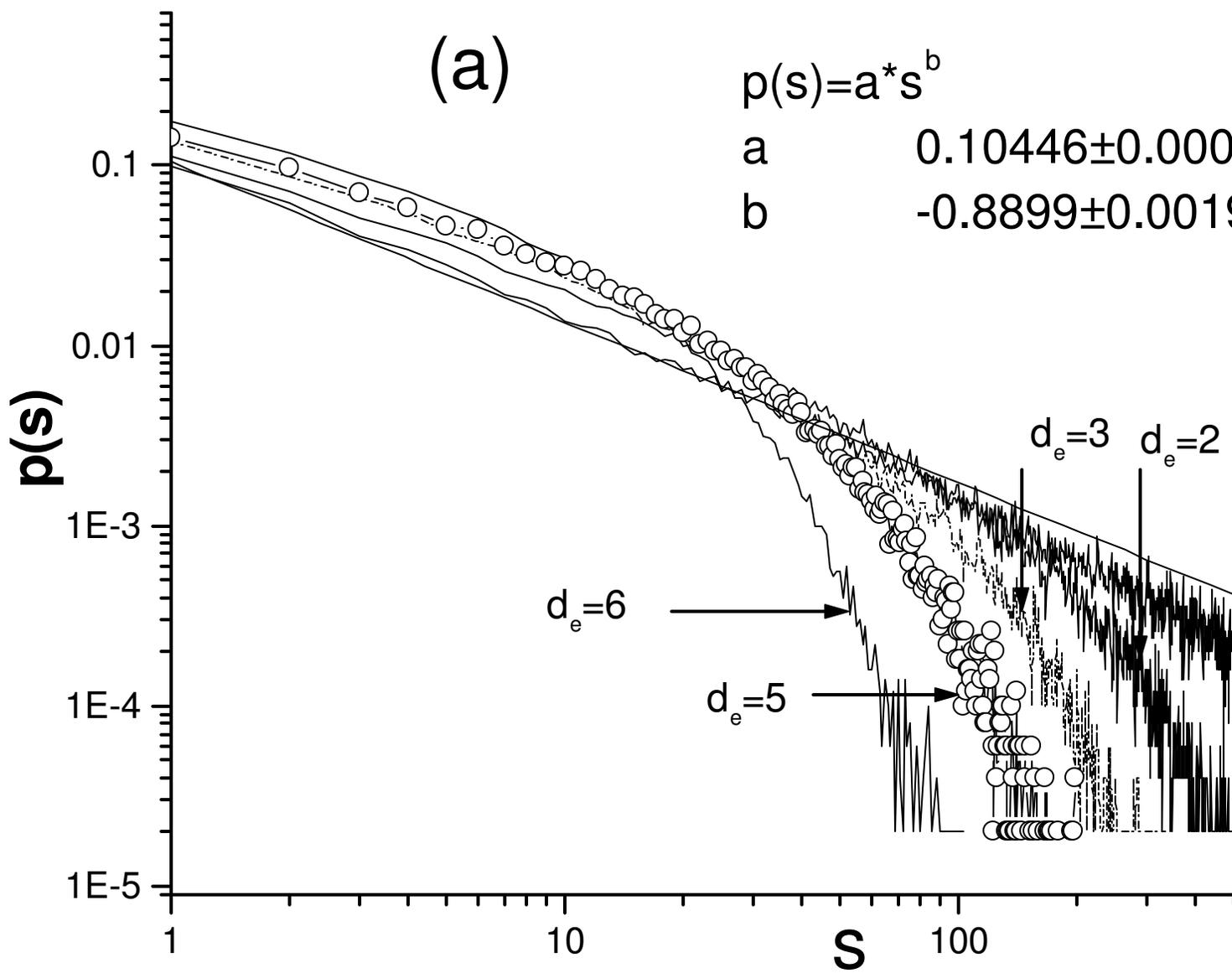
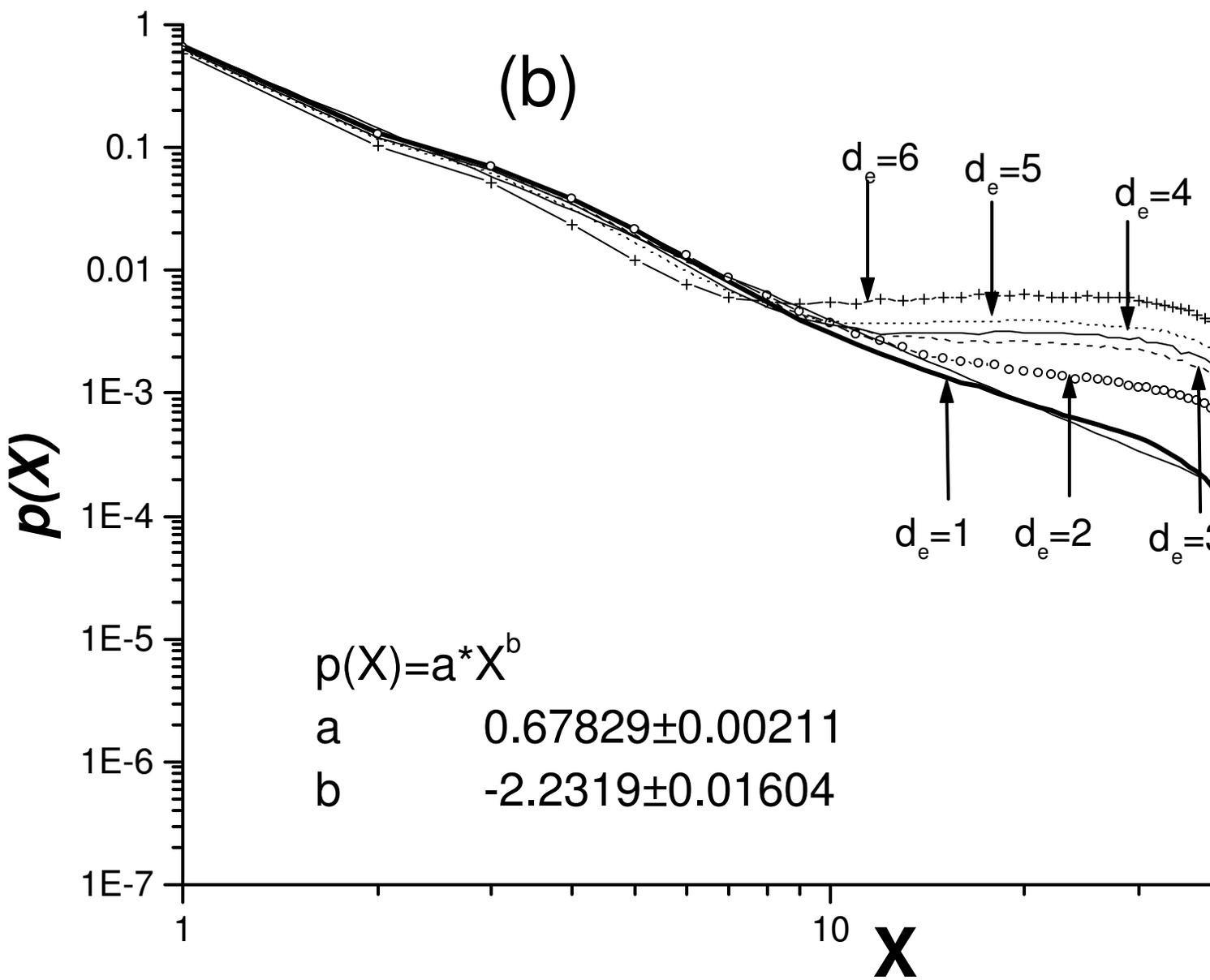
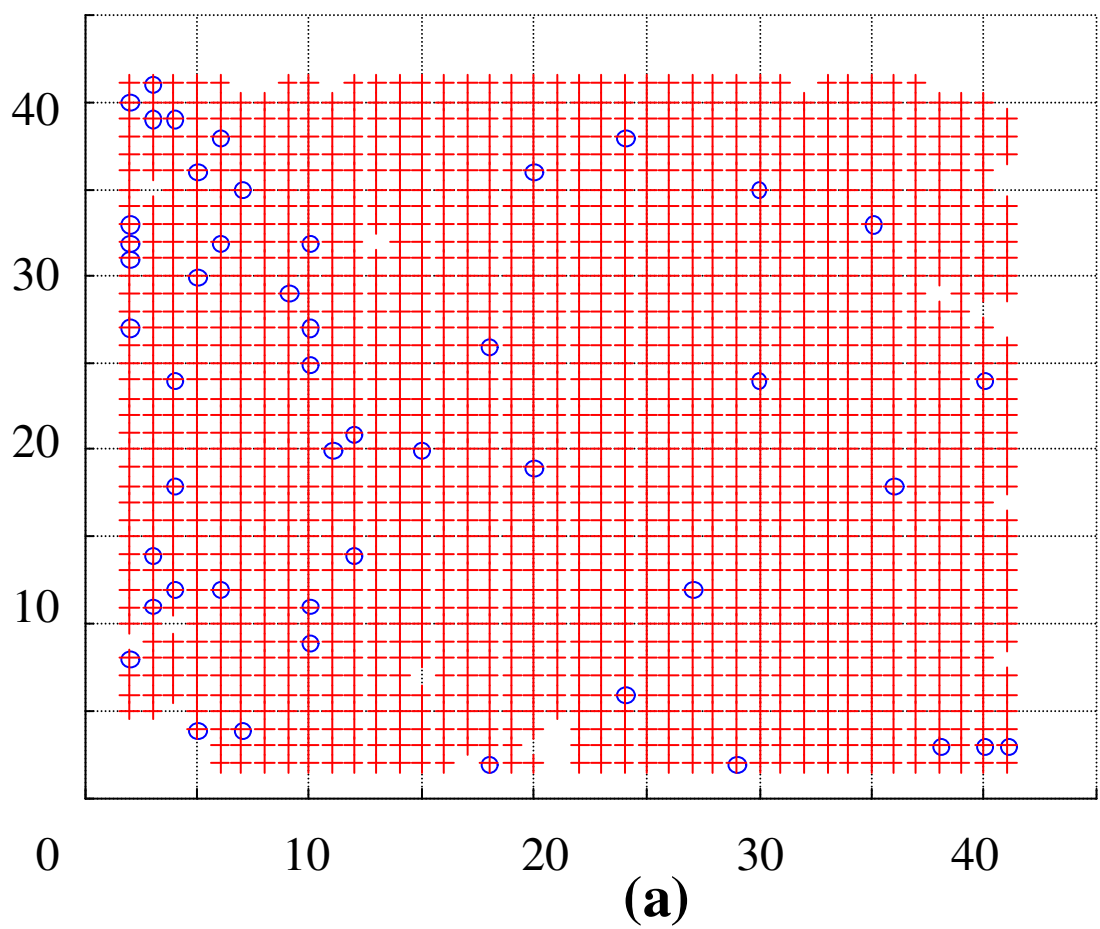
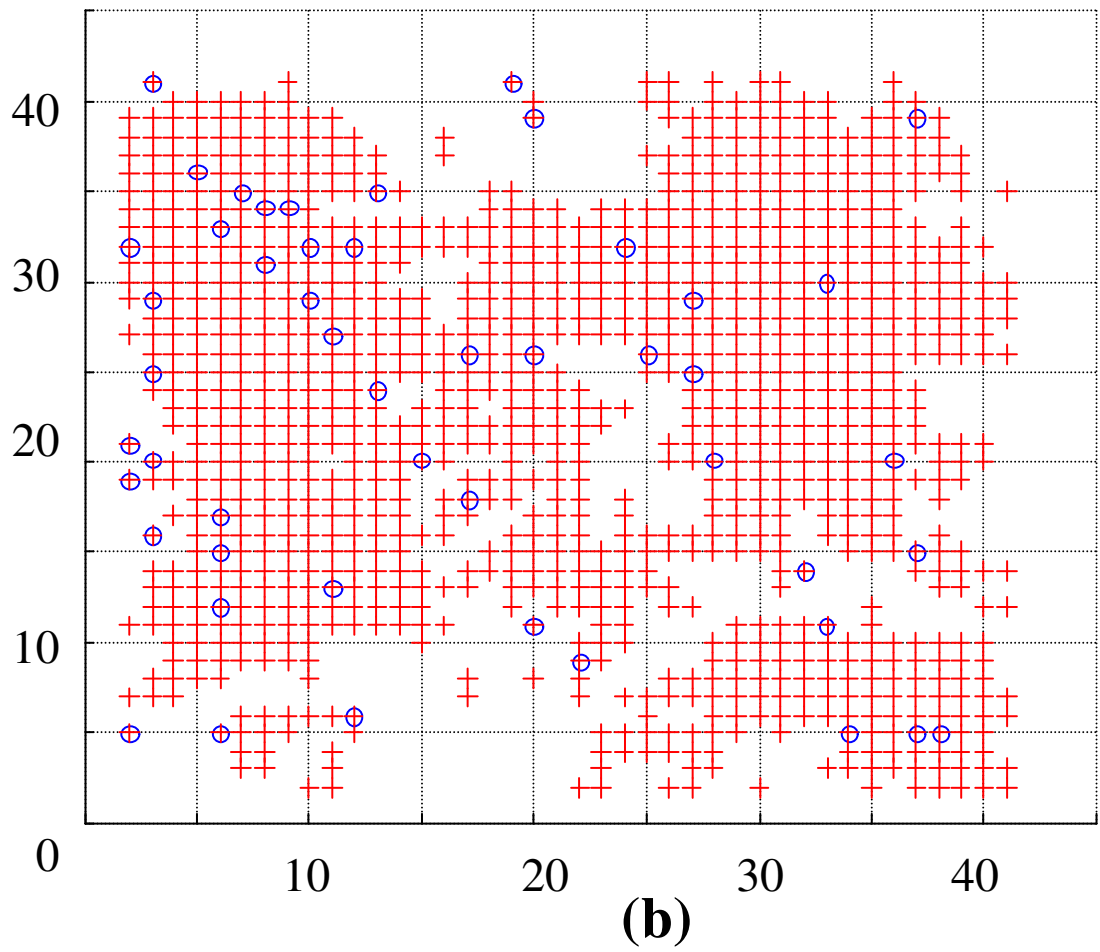
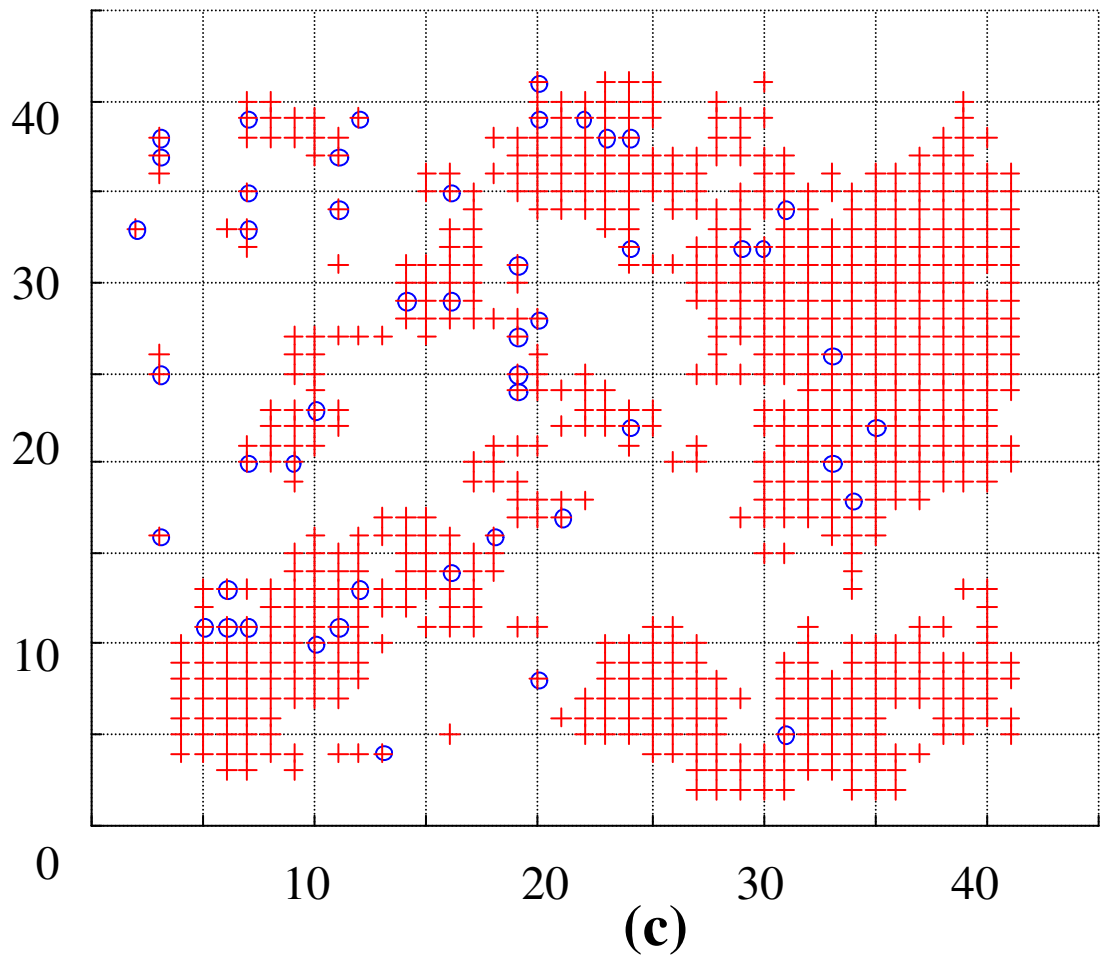
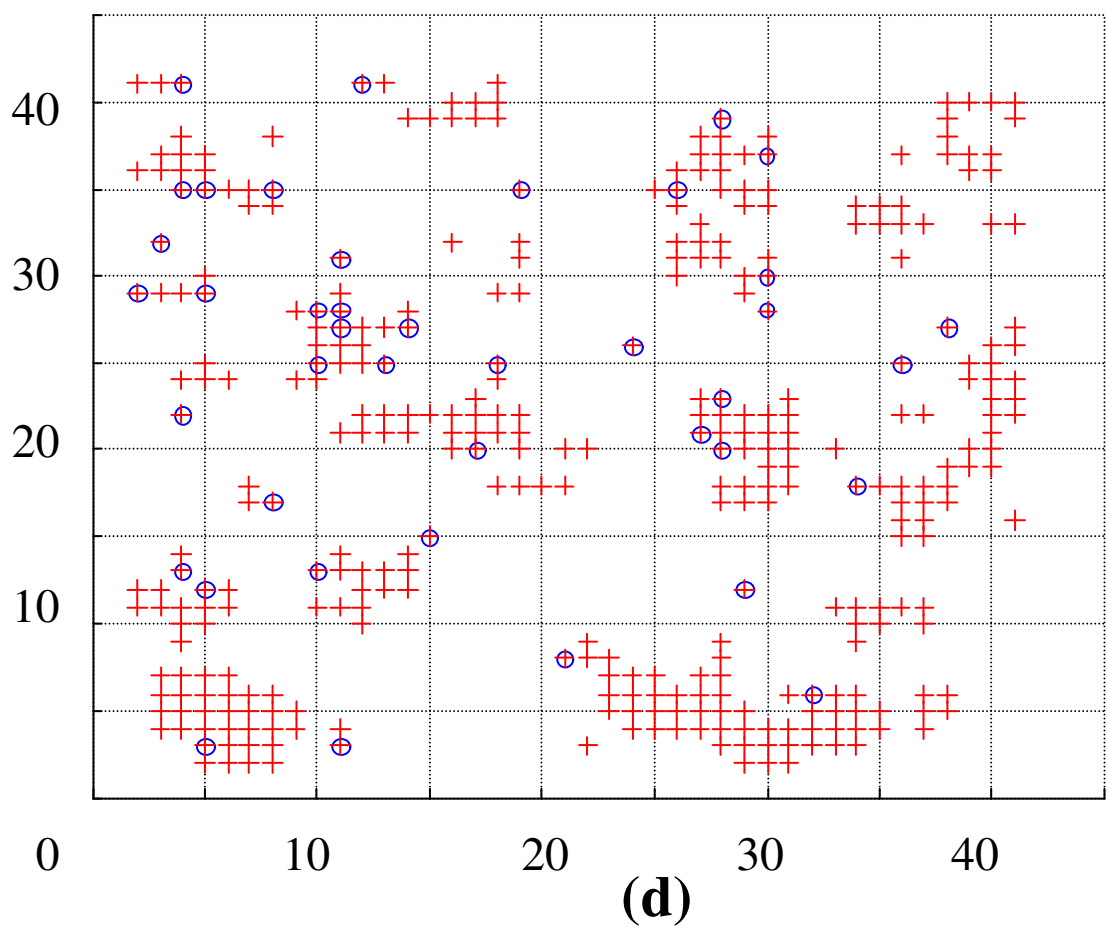
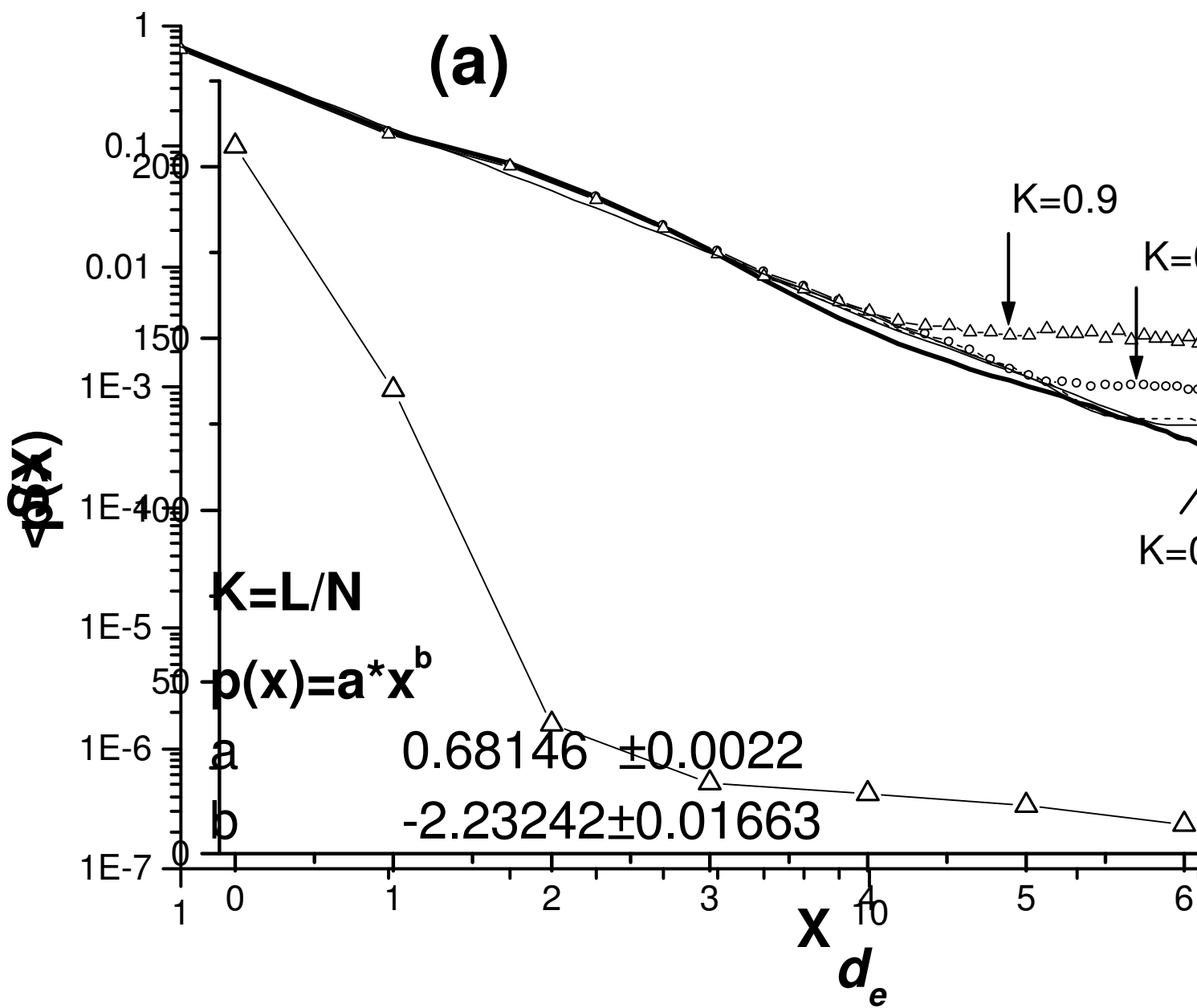
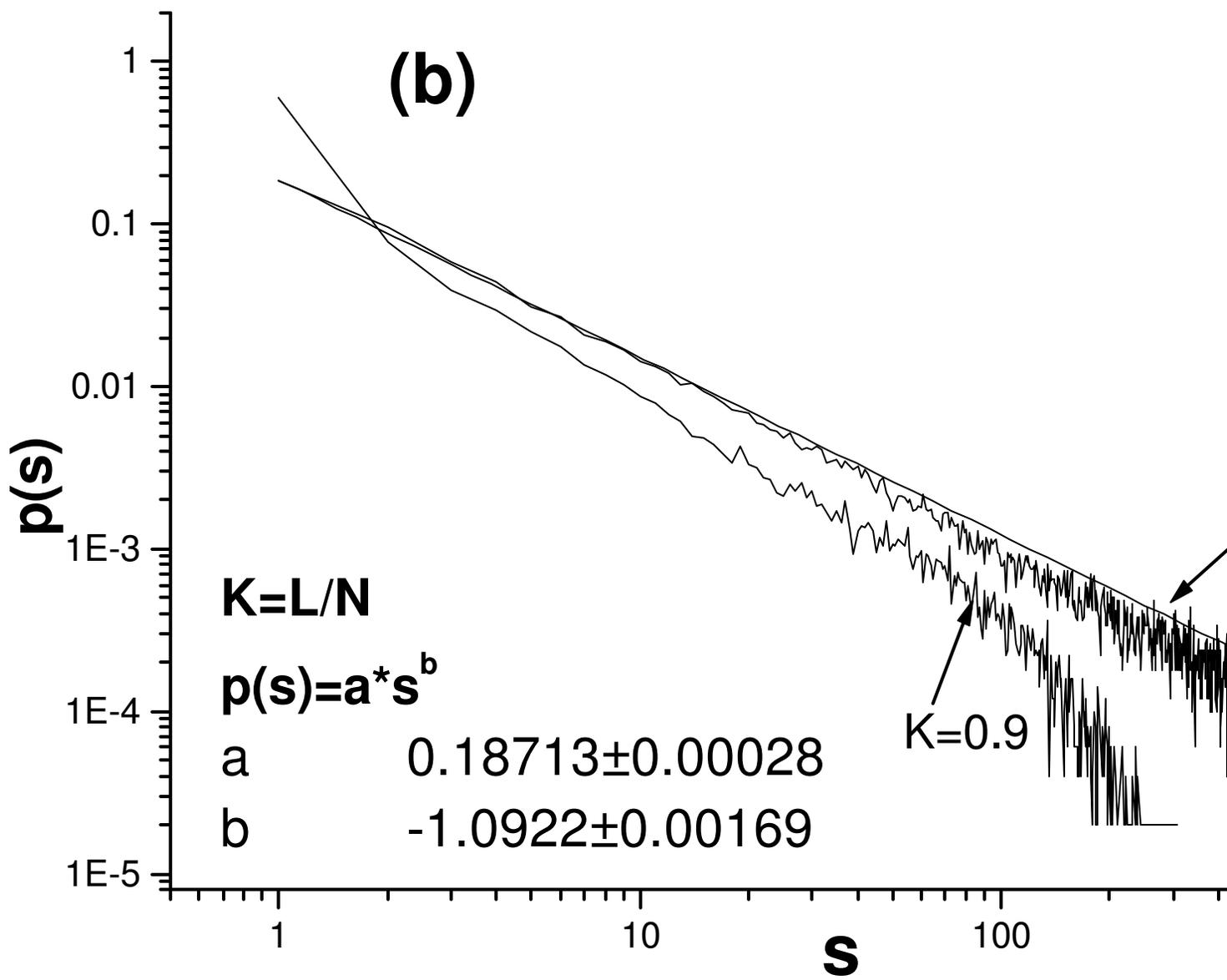
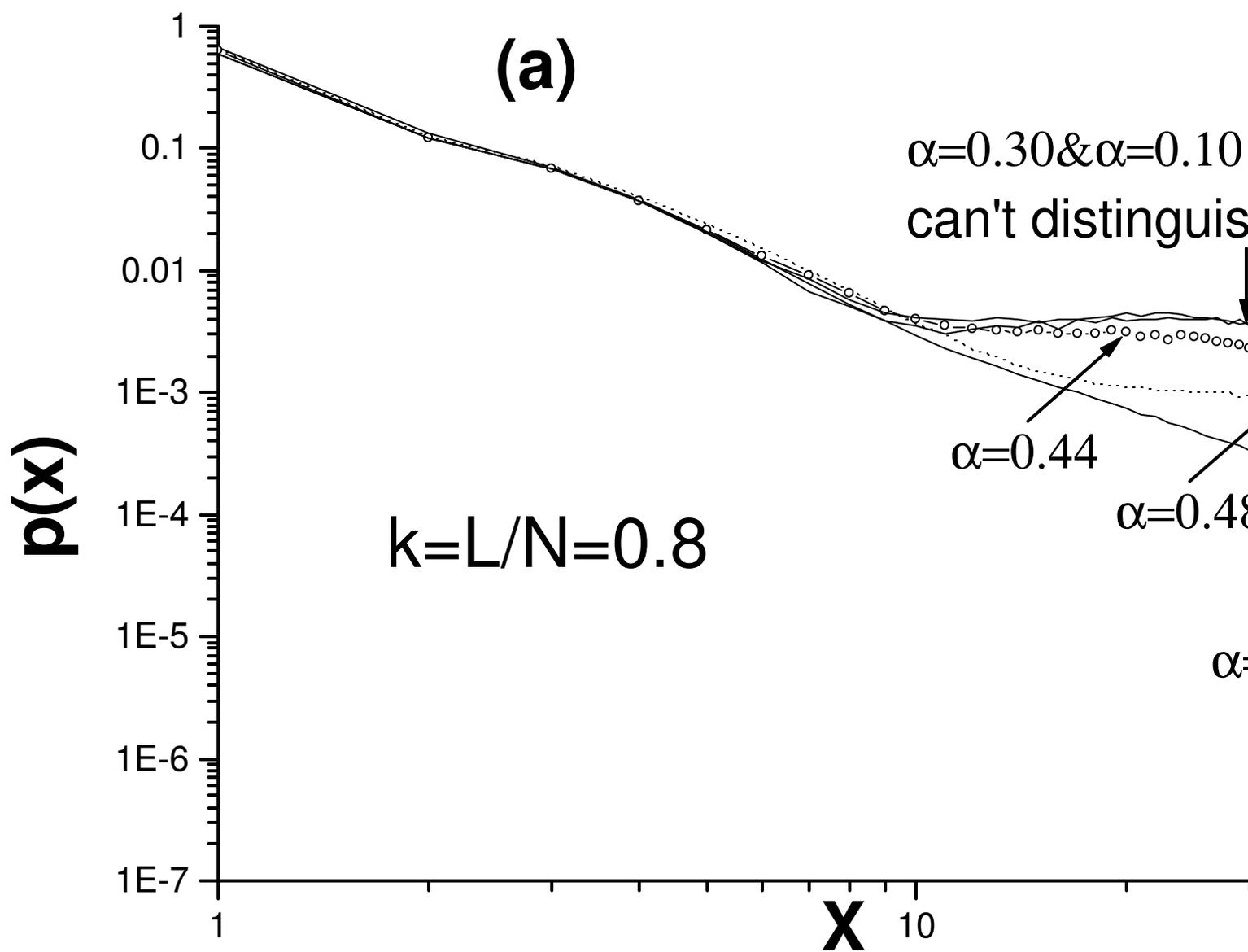
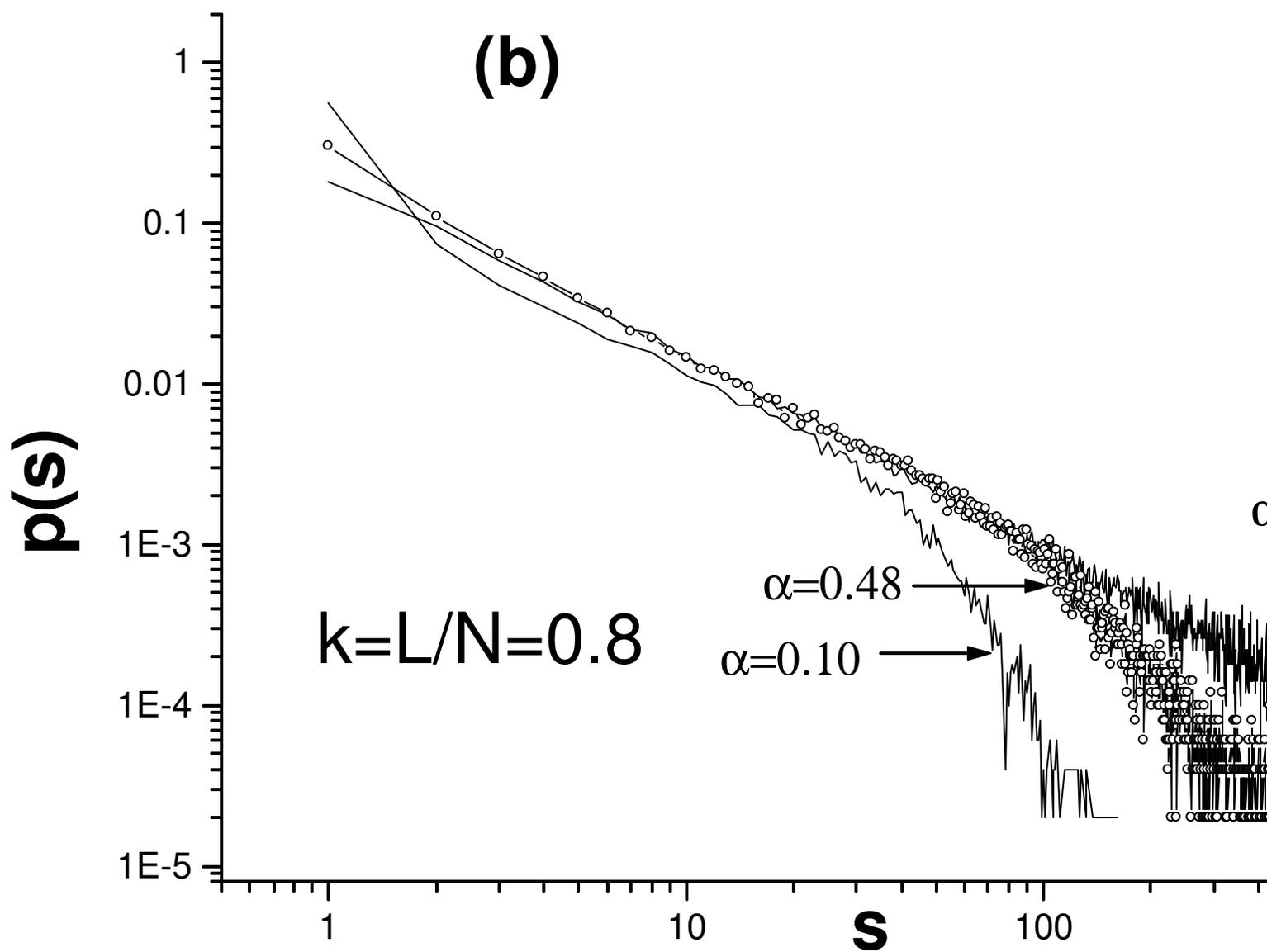
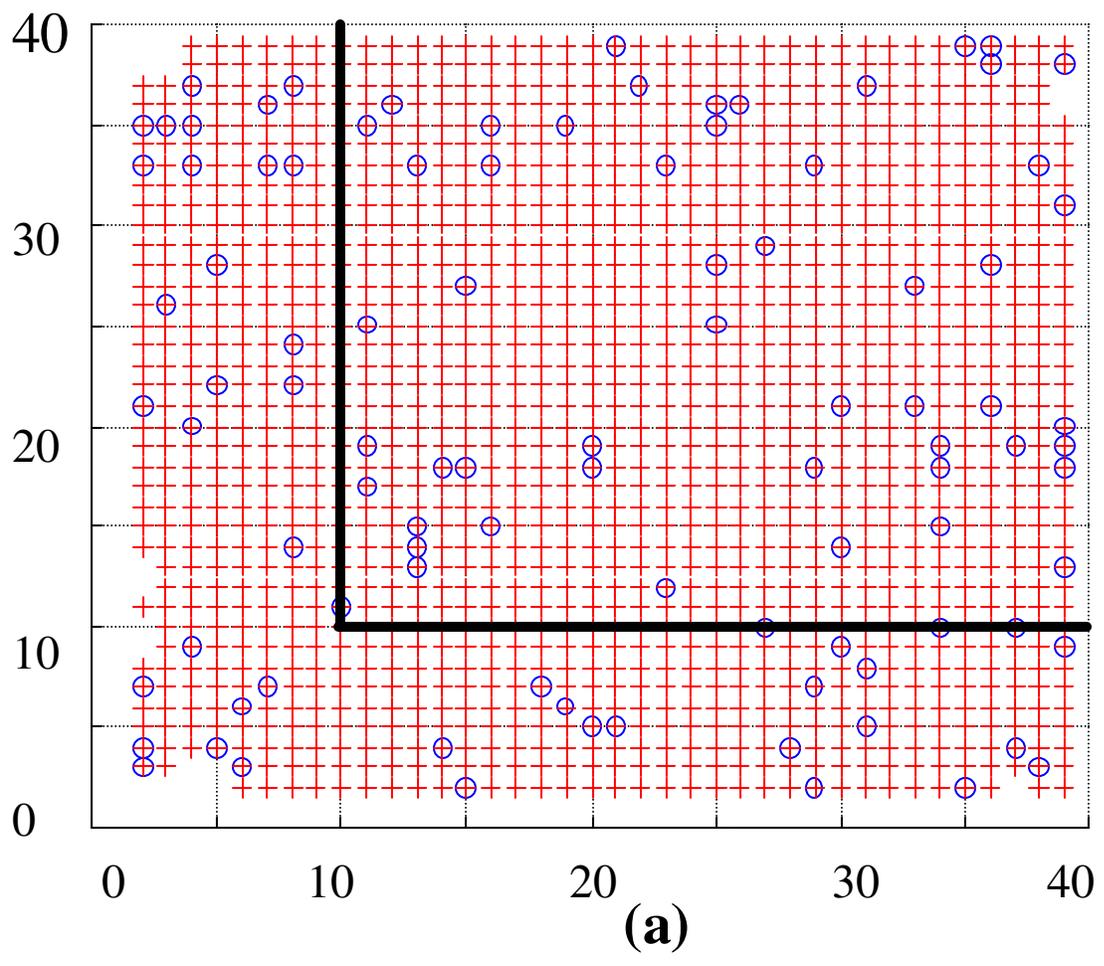
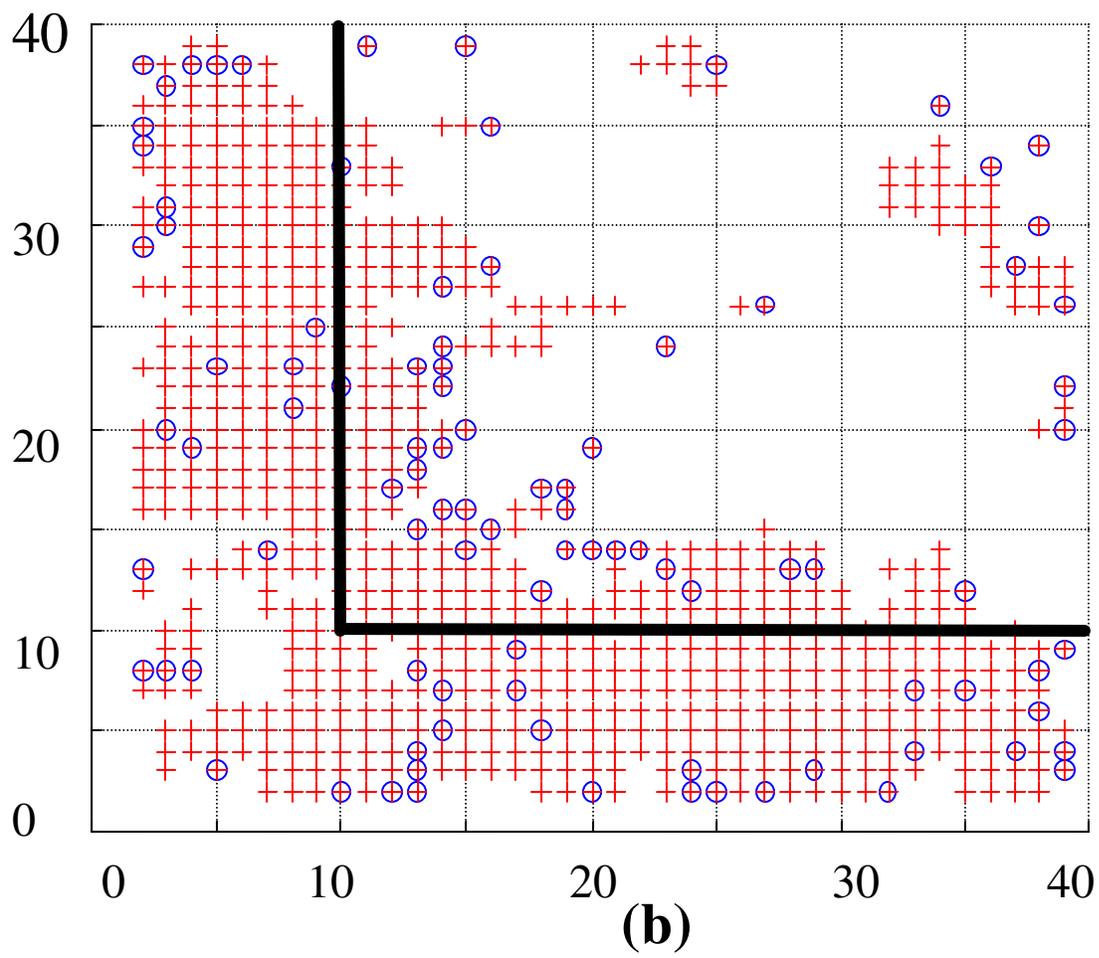
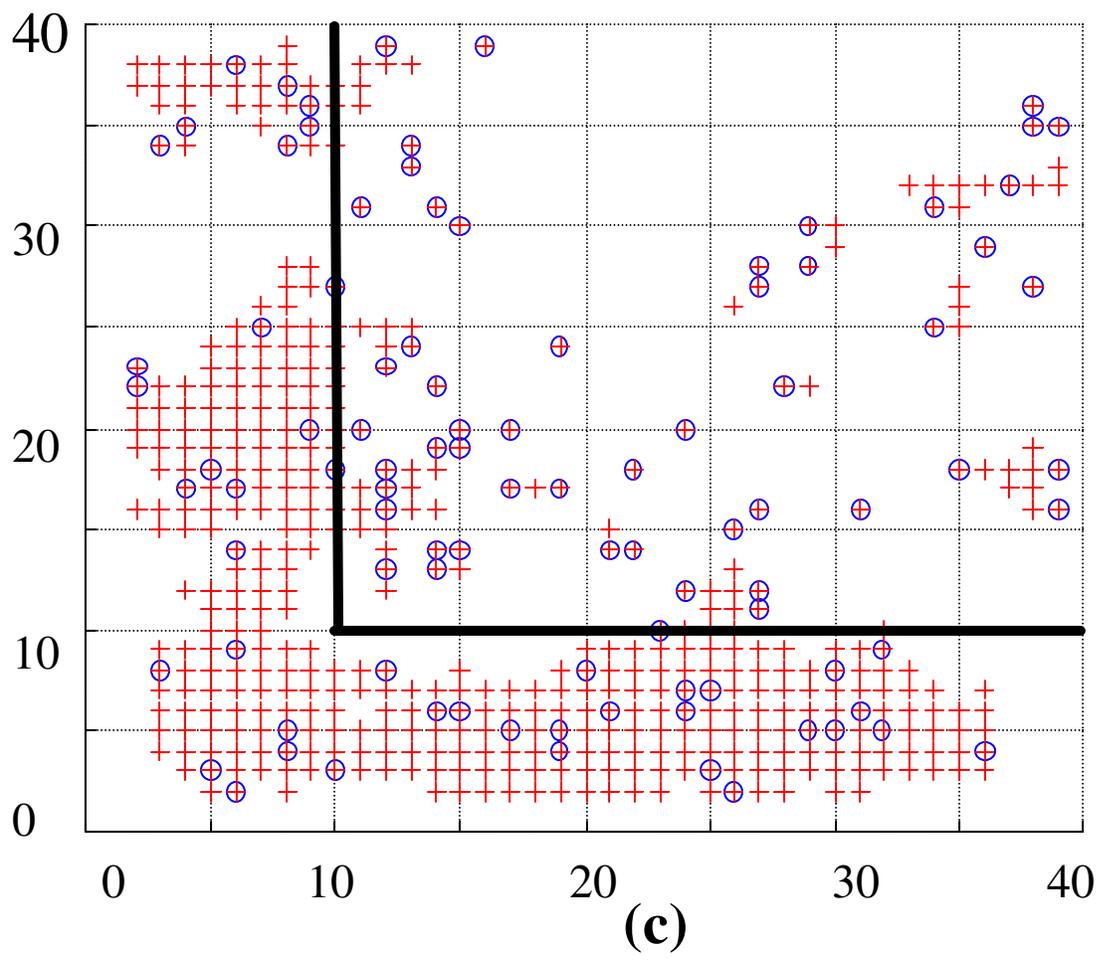
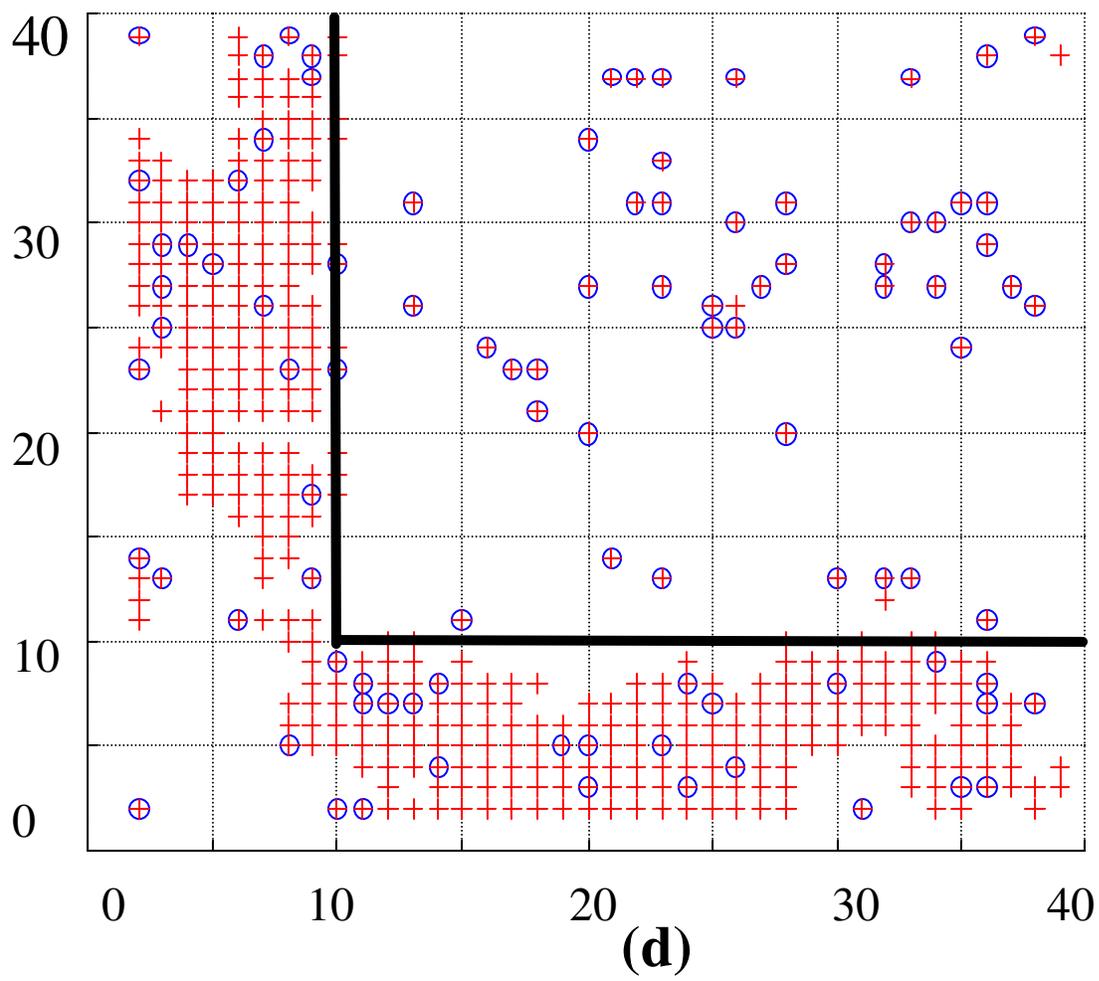
\end{document}